\documentstyle[amssymb,aps]{revtex}

\begin{document}
\title{Special reduction of density matrixes and entanglement between two bunches
of particles }
\author{Zai-Zhe Zhong}
\address{Department of Physics, Liaoning Normal University, Dalian 116029, Liaoning,\\
China. E-mail: zhongzaizheh@hotmail.com}
\maketitle

\begin{abstract}
By using of a special reduction way of density matrices, in this Letter we
find the entanglement between two bunches of particles, its measure can be
represented by the entanglement of formation.

PACC numbers: 03.65Ud, 03.67-a, 03.65Bz, 03.67Hk.
\end{abstract}

In this Letter, we discuss a special way of reduction of density matrices,
and prove that in an arbitrary multipartite qubit state there is the new
kind of entanglement, i.e. the entanglement between two bunches of
particles, independent what happens to the remaining particles, and its
measure can be represented by the entanglement of formation[1] $E_f$. Some
examples are discussed.

It is known that the problems of the measure of the entanglement of
bipartite qubit (pure- and mixed-)states are solved, e.g. see [1,2].
However, the description of entanglement of multipartite qubit states is a
formidable task as yet. Recently many new results have been obtained, one
way in which is to describe the properties of entanglement of multipartite
qubit states by using of the bipartite reduced density matrices and of the
entanglement of formation $E_f$. For instance, in [$3$] the entanglement
between two particles, independent what happens to the remaining particles,
is described by a bipartite reduced operator, and its measure is represented
by $E_f$. Some more new results concerned, see [4,5].

For the spin particle $M_k(k=1,2,\cdots ,N,\;N\geq 3),$ we simply write $%
\uparrow M_k>\equiv \mid 0>_k$ and $\downarrow M_k>\equiv \mid 1>_k$, or in
union by $\mid i>_k(i=0$ and $1).$ $\mid i>_k$spans the Hilbert space $H_k$.
The main points of the above ways are as follows: If $\rho $\ is a density
matrix acting upon $H_1\otimes H_2\otimes \cdots \otimes H_N,$\ let the $%
\rho _{kl}$\ be the bipartite reduced density matrix defined by 
\begin{equation}
\rho _{kl}\equiv tr_{1\cdots \stackrel{\times }{k}\cdots \stackrel{\times }{l%
}\cdots N}(\rho ),\;(1\leq k<l\leq N)
\end{equation}
where the $tr$\ denotes the trace of a matrix, and symbol $\times $\ denotes
the deletion operator. Since $\rho _{kl}$ is a bipartite density matrix, we
can use the entanglement of formation $E_f$. Therefore once use $\rho _{kl}$
and $E_f\left[ \rho _{kl}\right] $ to describe the entanglement status
between two particles $\mid i>_k$ and $\mid i>_l$, independent what happens
to the remaining particles.

About the above kind of ways by using of reduced density matrices $\rho
_{kl} $ and $E_f\left[ \rho _{kl}\right] $, we need to consider at least the
following two problems: First, why only consider we two particles, but not
more particles? In fact, for the case of entanglement among more particles
the problem backs again to the original status, i.e. we need to handle other
multipartite qubit state, the above way will runs up against difficulties.
For instance, when $N\geq 5$, and we need to consider how to describe the
entanglement among four particles $\mid i>_1,\mid i>_2,\mid i>_3$\ and $\mid
i>_4$, independent what happens to the remaining particles $\mid i>_5,\cdots
,\mid i>_N,$\ then we use the reduced matrix $\rho _{1234}\equiv tr_{5\cdots
N}(\rho ),$ however $\rho _{1234}$\ is not a bipartite qubit state, we
cannot use $E_f$. Although we still can write the set $\left\{ E_f\left[
\rho _{12}\right] ,E_f\left[ \rho _{13}\right] ,\cdots ,E_f\left[ \rho
_{34}\right] \right\} $, it cannot show more contents of above `entanglement
among the particles $\mid i>_1,\mid i>_2,\mid i>_3$\ and $\mid i>_4$', but
the latter must contain other more contents. How are we to surmount this
difficulty? Secondly, for some important multipartite qubit entangled
states, say $\mid $GHZ$_N>\equiv \frac 1{\sqrt{2}}\left( \mid 00\cdots
0>+\mid 11\cdots 1>\right) ,$\ $\rho _{GHZ_N}\equiv \mid $GHZ$_N><$GHZ$%
_N\mid ,$ all $\rho _{kl}$\ are disentangled, i.e. all $E_f\left[ \rho
_{kl}\right] =0,$\ this is somewhat making one puzzled: Is there some
possible entanglement shared between two parts of system $\mid $GHZ$_N>$\
with non-zero $E_f$? In this Letter, we suggest a way that ones should
consider some special ways of reduction of density matrices, then we prove
that there is a new kind of entanglement in multipartite qubit systems, i.e.
the entanglement between two bunches of particles, its measure can be
represented by $E_f.$ These results bring to light the more properties of
multipartite qubit entangled states.

In the first place, we discuss the simplest case, i.e. the tripartite qubit
entangled states. The general form of a $\Psi $ in $H_a\otimes H_b\otimes
H_c $ is as $\Psi =\sum_{i,j,k,=0,1}c_{ijk}\mid i>_a\otimes \mid j>_b\otimes
\mid k>_c(\;c_{ijk}\in {\Bbb C)}$. We define $H_{a/bc},H_{a/b\stackrel{\vee 
}{c}}$ , respectively by 
\begin{eqnarray}
H_{a/bc} &\equiv &\left\{ \Psi \mid _{:}\text{The form of }\Psi \text{ is as 
}\Psi _{a/bc}=\sum_{i,k=0,1}c_{ikk}\mid i>_a\otimes \mid k>_b\otimes \mid
k>_c\right\} \\
H_{a/b\stackrel{\vee }{c}} &\equiv &\left\{ \Psi \mid _{:}\text{The form of }%
\Psi \text{ is as }\Psi _{a/b\stackrel{\vee }{c}}=\sum_{i,k=0,1}c_{ik(1-k)}%
\mid i>_a\otimes \mid k>_b\otimes \mid 1-k>_c\right\}  \nonumber
\end{eqnarray}
$H_{a/bc},H_{a/b\stackrel{\vee }{c}}$ are two 4-dimensional subspaces
orthogonal to each other$,$ we have the direct sum decomposition 
\begin{equation}
H_a\otimes H_b\otimes H_c=H_{a/bc}\oplus H_{a/b\stackrel{\vee }{c}}
\end{equation}

Now we make two formal bases $\mid i>_x$ and $\mid i>_y$, since $H_{a/bc}$
and $H_{a/b\stackrel{\vee }{c}}$ both are 4-dimensional spaces, we take the
1-1 correspondences as 
\begin{equation}
\mid i>_a\otimes \mid k>_b\otimes k>_c\rightleftarrows \mid i>_a\otimes \mid
k>_x\text{and }\mid i>_a\otimes \mid k>_b\otimes \mid 1-k>_c\rightleftarrows
\mid i>_a\otimes \mid k>_y
\end{equation}
then we have the following isomorphisms 
\begin{equation}
H_{a/bc}\approx H_a\otimes H_x\text{, }\;H_{a/b\stackrel{\vee }{c}}\approx
H_a\otimes H_y,\;H_a\otimes H_b\otimes H_c\approx H_a\otimes H_x\oplus
H_a\otimes H_y
\end{equation}

From Eq.(5), any quantum state $\Psi $ can be expressed in only one form as
a sum of two orthogonal states $\Psi _{a/bc}$ and $\Psi _{a/b\stackrel{\vee 
}{c}},$ especially, we can explain $\Psi _{a/bc}$ and $\Psi _{a/b\stackrel{%
\vee }{c}}$ as follows: $\Psi _{a/bc}$ is the `wave function describing two
bunches ($\mid i>_a)$ and $\left( \mid k>_b\otimes \mid k>_c\right) $'$,$
where in bunch $\left( \mid k>_b\otimes \mid k>_c\right) $ the
spin-directions of particles $b$ and $c$ always are the same, hence their
behavior of spin can be regarded, as a whole, like to a single spin
particle. $\Psi _{a/b\stackrel{\vee }{c}}$ is the `wave function describing
two bunches ($\mid i>_a)$ and $\left( \mid k>_b\otimes \mid 1-k>_c\right) $'$%
,$ where in bunch $\left( \mid k>_b\otimes \mid 1-k>_c\right) $ the spin
directions of particles $b$ and $c$ always are contrary, hence their
behavior also can be regarded, as a whole, like to other single spin
particle.

The projection from $H_a\otimes H_b\otimes H_c$ to $H_a\otimes H_x$ and $%
H_a\otimes H_y$, respectively, are two left multiplication operators as 
\begin{eqnarray}
P_{a/bc} &\equiv &\sum_{i,k=0,1}\mid i>_a\otimes \mid k>_x<_ck\mid \otimes
<_bk\mid \otimes <_ai\mid :\Psi \longrightarrow \Psi _{ax}\equiv
P_{a/bc}\left( \Psi \right) \;=\sum_{i,k=0,1}c_{ikk}\mid i>_a\otimes \mid
k>_x \\
P_{a/b\stackrel{\vee }{c}} &\equiv &\sum_{i,k=0,1}\mid i>_a\otimes \mid
k>_x<_c1-k\mid \otimes <_bk\mid \otimes <_ai\mid :\Psi \longrightarrow \Psi
_{ay}\equiv P_{a/b\stackrel{\vee }{c}}\left( \Psi \right)
\;=\sum_{i,k=0,1}c_{ik\left( 1-k\right) }\mid i>_a\otimes \mid k>_y 
\nonumber
\end{eqnarray}
And the interior mappings $I_{ax}:H_a\otimes H_x$ $\longrightarrow $ $%
H_a\otimes H_b\otimes H_c$ and $I_{ay}:\;H_a\otimes H_y{}$ $\longrightarrow $
$H_a\otimes H_b\otimes H_c$ are two left multiplication operators as 
\begin{eqnarray}
I_{ax} &\equiv &\sum_{i,k=0,1}\mid i>_a\otimes \mid k>_b\otimes \mid
k>_c<_xk\mid \otimes <_ai\mid :\Psi _{ax}\longrightarrow I_{ax}\left( \Psi
_{ax}\right) =\Psi _{a/bc}. \\
I_{ay} &\equiv &\sum_{i,k=0,1}\mid i>_a\otimes \mid k>_b\otimes \mid
1-k>_c<_xk\mid \otimes <_ai\mid :\Psi _{ay}\longrightarrow I_{ax}\left( \Psi
_{ax}\right) =\Psi _{a/b\stackrel{\vee }{c}}.  \nonumber
\end{eqnarray}
Obviously, $P_{a/bc}\circ I_{ax}$ and $P_{a/b\stackrel{\vee }{c}}\circ
I_{ay} $, respectively, are the identical mappings upon $H_x$ and $H_y$.

Suppose that $T\;$is an linear operator, 
\begin{equation}
T:\;H_a\otimes H_b\otimes H_c\longrightarrow H_a\otimes H_b\otimes H_c,\;%
\widetilde{\Psi }=T\left( \Psi \right) 
\end{equation}
then we can obtain the induced mappings $T_{ax}$ and $T_{ay}$ from $T$, 
\begin{eqnarray}
T_{ax} &:&\;\;H_a\otimes H_x\longrightarrow \;H_a\otimes H_x\text{, }\Psi
_{ax}\longrightarrow \;\widetilde{\Psi }_{ax}\equiv T_{ax}\left( \Psi
_{ax}\right) =P_{a/bc}\circ T\circ I_{ax}\left( \Psi _{ax}\right)   \nonumber
\\
T_{ay} &:&\;\;H_a\otimes H_y\longrightarrow \;H_a\otimes H_y\text{, }\Psi
_{ay}\longrightarrow \;\widetilde{\Psi }_{ax}\equiv T_{ay}\left( \Psi
_{ay}\right) =P_{a/b\stackrel{\vee }{c}}\circ T\circ I_{ay}\left( \Psi
_{ay}\right) 
\end{eqnarray}
We take especially an interest in the case of that $T$ is a ( pure or mixed)
density operator $\rho $ on $H_a\otimes H_b\otimes H_c,$ in this case the
results obtained are two bipartite operators $\Upsilon _{axd}$ and $\Upsilon
_{ayd}$ as 
\begin{eqnarray}
\Upsilon _{ax} &:&\Psi _{axd}\longrightarrow \;\Psi _{ax}^{\prime }\equiv
\Upsilon _{ax}\left( \Psi _{ax}\right) ,\;\Upsilon _{ax}\equiv P_{a/bc}\circ
\rho \circ I_{ax}  \nonumber \\
\Upsilon _{ay} &:&\Psi _{ayd}\longrightarrow \;\Psi _{ay}^{\prime }\equiv
\Upsilon _{ay}\left( \Psi _{ay}\right) ,\;\Upsilon _{ay}\equiv P_{a/b%
\stackrel{\vee }{c}}\circ \rho \circ I_{ay}
\end{eqnarray}
By using of Eqs.($6),(7)$ and (9), we find the entries of $\Upsilon _{ax}$
and $\Upsilon _{ay}$, respectively, are 
\begin{equation}
\left[ \Upsilon _{ax}\right] _{ij,kl}=\left[ \rho \right]
_{ijj,kll},\;\left[ \Upsilon _{ay}\right] _{ij,kl}=\left[ \rho \right]
_{ij(1-j),kl(1-l)},(i,j,k,l=0,1)
\end{equation}
where $\left[ \rho \right] _{ijm,kls}$ are the entries of density matrix $%
\rho $. By normalization, we can write 
\begin{equation}
\Upsilon _{ax}=\eta _{ax}\rho _{ax}\text{, }\eta _{ax}\equiv
\sum_{r,s=0,1}\left[ \rho \right] _{rss,rss},\;\rho _{ax}\equiv \frac 1{\eta
_{ax}}\Upsilon _{ax}.\;\Upsilon _{ay}=\eta _{ay}\rho _{ay}\text{, }\eta
_{ay}\equiv \sum_{r,s=0,1}\left[ \rho \right] _{rs(1-s),rs\left( 1-s\right)
},\;\rho _{ay}\equiv \frac 1{\eta _{ay}}\Upsilon _{ay}
\end{equation}
Since $\rho $ is a density matrix, from Eqs.(10),(11) and (12) we can
directly verify that $\rho _{ax}$\ and $\rho _{ay}$\ both are bipartite
density operators. Obviously, $\rho _{ax}$\ describes the status of
entanglement between bunches $\left( \mid i>_a\right) $ and $\left( \mid
k>_b\otimes \mid k>_c\right) ,$ and $\rho _{ay}$ describes the status of
entanglement between bunches bunches $\left( \mid i>_a\right) $ and $\left(
\mid k>_b\otimes \mid 1-k>_c\right) .$ In addition, there is the relation%
{\bf \ }$\eta _{ax}+\eta _{ay}=1.$ This means that we can consider the
operator $\rho _{\left( a,bc\right) }$ defined by 
\begin{eqnarray}
\rho _{\left( a,bc\right) } &\equiv &\Upsilon _{ax}+\Upsilon _{ay}=\eta
_{ax}\rho _{ax}+\eta _{ay}\rho _{ay}  \nonumber \\
&=&\left[ 
\begin{array}{cccc}
\left[ \rho \right] _{000,000}+\left[ \rho \right] _{001,001} & \left[ \rho
\right] _{000,011}+\left[ \rho \right] _{001,010} & \left[ \rho \right]
_{000,100}+\left[ \rho \right] _{001,101} & \left[ \rho \right]
_{000,111}+\left[ \rho \right] _{001,110} \\ 
\left[ \rho \right] _{011,000}+\left[ \rho \right] _{010,001} & \left[ \rho
\right] _{011,011}+\left[ \rho \right] _{010,010} & \left[ \rho \right]
_{011,100}+\left[ \rho \right] _{010,101} & \left[ \rho \right]
_{011,111}+\left[ \rho \right] _{010,110} \\ 
\left[ \rho \right] _{100,000}+\left[ \rho \right] _{101,001} & \left[ \rho
\right] _{100,011}+\left[ \rho \right] _{101,010} & \left[ \rho \right]
_{100,100}+\left[ \rho \right] _{101,101} & \left[ \rho \right]
_{100,111}+\left[ \rho \right] _{101,110} \\ 
\left[ \rho \right] _{111,000}+\left[ \rho \right] _{110,001} & \left[ \rho
\right] _{111,011}+\left[ \rho \right] _{110,001} & \left[ \rho \right]
_{111,100}+\left[ \rho \right] _{110,101} & \left[ \rho \right]
_{111,111}+\left[ \rho \right] _{110,110}
\end{array}
\right] 
\end{eqnarray}
{\bf then }$\rho _{\left( a,bc\right) }${\bf \ can be taken as a bipartite
qubit mixed-state, which describes the status of entanglement between two
bunches of particles }$\left( a\right) ${\bf \ and }$\left( b,c\right) ${\bf %
. In addition, the procedure in accordance with the rules in Eqs.(11), (12)
and (13), in fact, is a special reduction of density matrices.}

Similarly, we take 
\begin{eqnarray}
H_{b/ca} &\equiv &\left\{ \Psi \mid _{:}\text{The form of }\Psi \text{ is as 
}\Psi _{b/ca}=\sum_{i,k=0,1}c_{kik}\mid k>_a\otimes \mid i>_b\otimes \mid
k>_c\right\}  \nonumber \\
H_{b/c\stackrel{\vee }{a}} &\equiv &\left\{ \Psi \mid _{:}\text{The form of }%
\Psi \text{ is as }\Psi _{b/c\stackrel{\vee }{a}}=\sum_{i,k=0,1}c_{(1-k)ik}%
\mid 1-k>_a\otimes \mid i>_b\otimes \mid k>_c\right\} \\
\text{Correspoces} &\mid &k>_a\otimes \mid i>_b\otimes \mid
k>_c\rightleftarrows \mid i>_b\otimes \mid k>_x\Longrightarrow
H_{b/ca}\approx H_b\otimes H_x  \nonumber \\
&\mid &1-k>_a\otimes \mid i>_b\otimes \mid k>_c\rightleftarrows \mid
i>_b\otimes \mid k>_y\Longrightarrow H_{b/c\stackrel{\vee }{a}}\text{ }%
\approx H_b\otimes H_y  \nonumber
\end{eqnarray}
and similarly construct the projections $P_{b/ca},P_{b/c\stackrel{\vee }{a}%
}, $ the interior mappings $I_{b/ca}$, $I_{b/c\stackrel{\vee }{a}}$ and the
induced mappings $\Upsilon _{bx},\Upsilon _{by},\cdots ,$etc.. They lead to 
\begin{eqnarray}
\rho _{(b,ca)} &\equiv &\Upsilon _{bx}+\Upsilon _{by},\;\left[ \rho
_{(b,ca)}\right] _{ij,kl}\equiv \left[ \rho \right] _{jij,lkl}+\left[ \rho
\right] _{(1-j)ij,(1-l)kl}\; \\
\rho _{(b,ca)} &=&\left[ 
\begin{array}{cccc}
\rho _{000,000}+\rho _{100,100} & \rho _{000,101}+\rho _{100,001} & \rho
_{000,010}+\rho _{100,110} & \rho _{000,111}+\rho _{100,011} \\ 
\rho _{101,000}+\rho _{001,100} & \rho _{101,101}+\rho _{001,001} & \rho
_{101,010}+\rho _{001,110} & \rho _{101,111}+\rho _{001,011} \\ 
\rho _{010,000}+\rho _{110,100} & \rho _{010,101}+\rho _{110,001} & \rho
_{010,010}+\rho _{110,110} & \rho _{010,111}+\rho _{110,011} \\ 
\rho _{111,000}+\rho _{011,100} & \rho _{111,101}+\rho _{011,001} & \rho
_{111,010}+\rho _{011,110} & \rho _{111,111}+\rho _{011,011}
\end{array}
\right]  \nonumber
\end{eqnarray}
$\rho _{(b,ca)}$ is a bipartite density matrix which describes the status of
entanglement shared between two bunches of particles $\left( b\right) $\ and%
{\bf \ }$\left( c,a\right) $. Similarly, we can yet obtain 
\begin{equation}
\rho _{(c,ab)}\equiv \Upsilon _{cx}+\Upsilon _{cy},\;\left[ \rho
_{(c,ab)}\right] _{ij,kl}=\left[ \rho \right] _{jji,llk}+\left[ \rho \right]
_{j(1-j)i,l(1-l)k}
\end{equation}
Notice that although we can yet write $\rho _{\left( ab,c\right) },\cdots ,$
there are repeats, e.g. $\rho _{\left( ab,c\right) }=\rho _{(c,ab)},\cdots ,$
etc., there only are three independent $\rho _{\left( \bullet ,\bullet
\bullet \right) }$, i.e. $\rho _{\left( a,bc\right) },$ $\rho _{\left(
b,ca\right) }$ and $\rho _{\left( c,ab\right) }.$

Since $\rho _{\left( a,bc\right) },$ $\rho _{\left( b,ca\right) }$ and $\rho
_{\left( c,ab\right) }$ all are bipartite density matrix, we naturally use $%
E_f$ to represent their entanglement measure. For a given $\rho $ this
measure $E_f$ can be concretely calculated by using of the so-called
`concurrence'[$6,7]$. For instance, for $\rho _{\left( a,bc\right) }$
defined as in Eq.(13) 
\begin{equation}
E_f\left[ \rho _{\left( a,bc\right) }\right] =h\left( \frac 12+\frac 12\sqrt{%
1-C^2}\right)
\end{equation}
where $h$ is the binary entropy function $h\left( x\right) \equiv -x\log
_2x-\left( 1-x\right) \log _2\left( 1-x\right) ,$ the concurrence $C$ is
determined by 
\begin{equation}
C=\max \left\{ 0,-\lambda _1,-\lambda _2,-\lambda _3,-\lambda _4,\right\}
\end{equation}
where $\lambda _\iota $ are the eigenvalues, in decreasing order, of the
Hermitian matrix $R\equiv \sqrt{\sqrt{\rho _{\left( a,bc\right) }}\stackrel{%
\thicksim }{\rho _{\left( a,bc\right) }}\sqrt{\rho _{\left( a,bc\right) }}}%
,\;\stackrel{\thicksim }{\rho _{\left( a,bc\right) }}=\left( \sigma
_2\otimes \sigma _2\right) \left( \rho _{\left( a,bc\right) }\right)
^{*}\left( \sigma _2\otimes \sigma _2\right) ,\;\sigma _2$ is the Pauli
matrix $\left[ 
\begin{array}{cc}
0 & -i \\ 
i & 0
\end{array}
\right] ,$ the star is the complex conjugation.. Similarly, for $\rho
_{\left( b,ca\right) }$ and $\rho _{\left( c,ab\right) }.$ Therefore, we can
obtain the complete set of measures of entanglement shared between every
pair of two bunches, i.e. $\left\{ E_f\left[ \rho _{(a,bc)}\right]
,E_f\left[ \rho _{(b,ca)}\right] ,E_f\left[ \rho _{(c,ab)}\right] \right\} $%
, it describes some character of the entanglement status among three
particles $a$, $b,$ and $c$.

Now, we return to the problems mentioned in the start of this Letter. In the
first place, if $\rho $ is a density operator upon $H_1\otimes H_2\otimes
\cdots \otimes H_N$ $(N\geq 3),$ then $tr_{1\cdots \stackrel{\times }{j}%
\cdots \stackrel{\times }{k}\cdots \stackrel{\times }{l}\cdots N}\left( \rho
\right) $ ($1\leq j<k<l\leq N)$ is a tripartite density operator acting upon 
$H_i\otimes H_k\otimes H_l,$ {\bf therefore the status of the entanglement
between two bunches of particles }$\left( M_j\right) ${\bf \ and }$\left(
M_k,M_l\right) ${\bf , independent what happens to the remaining particles,
can be described by state }$\left[ tr_{1\cdots \stackrel{\times }{j}\cdots 
\stackrel{\times }{k}\cdots \stackrel{\times }{l}\cdots N}\left( \rho
\right) \right] _{\left( j,kl\right) }.$ In the following we simply write 
\begin{equation}
\rho _{(j,kl)}\equiv \left[ tr_{1\cdots \stackrel{\times }{j}\cdots 
\stackrel{\times }{k}\cdots \stackrel{\times }{l}\cdots N}\left( \rho
\right) \right] _{\left( j,kl\right) }
\end{equation}
Here we must stress that $\rho _{(j,kl)}${\bf \ is a special reduced matrix
of }$\rho ${\bf \ by two reduction procedures in succession: The first is
the ordinary reduction, the second is in accordance with the special rules
as in Eqs.(11), (12) and (13).} The measure $E_f\left[ \rho _{(j,kl)}\right] 
$ can be calculated as in Eqs.(17) and (18). Similarly, for $\rho _{(k,lj)}$
and $\rho _{(l,jk)}.$ For three particles $\mid i>_j,\mid i>_k,\mid i>_l,$%
the set $\left\{ E_f\left[ \rho _{(j,kl)}\right] ,E_f\left[ \rho
_{(k,lj)}\right] ,E_f\left[ \rho _{(l,jk)}\right] \right\} $ completely
describes all entanglement between every pair consisting of a bunch
containing single particle and a bunch containing two particles . Obviously,
this shows a character of the entanglement among three particles $\mid
i>_j,\mid i>_k,\mid i>_l$ in the multipartite qubit state $\rho $,
independent what happens to the remaining particles. This cannot be obtained
only by using of the ordinary reduction as in Eq.($1$) and $E_f$.

The generalization of more high dimensional bunches is straightforward, e.g.
we can obtain $\rho _{(12,34)}(N\geq 4)$ from $\rho _{12/34},\rho _{1%
\stackrel{\vee }{2}/34},\rho _{12/3\stackrel{\vee }{4}},\rho _{1\stackrel{%
\vee }{2}/3\stackrel{\vee }{4}},$ and obtain $\rho _{(1,234)}$ from $\rho
_{1/234},\rho _{1/2\stackrel{\vee }{3}4},\rho _{1/23\stackrel{\vee }{4}%
},,\rho _{1/2\stackrel{\vee }{3}\stackrel{\vee }{4}}$\ (notice that, in
fact, $\rho _{1/\stackrel{\vee }{2}3\stackrel{\vee }{4}}=\rho _{1/2\stackrel{%
\vee }{3}4}$, etc.). Similarly, $\rho _{\left( 13,24\right) }$ ,$\rho
_{\left( 14,23\right) },\cdots $, etc.. At last, when N particles are
divided into two bunches $\left\{ r_1,\cdots ,r_m\right\} $ and $\left\{
s_1,\cdots ,s_n\right\} ,$ where $1\leq r_1<r_2<\cdots <r_m\leq N,\;1\leq
s_1<s_2<\cdots <s_n\leq N,\left\{ r_1,\cdots ,r_m\right\} \cap \left\{
s_1,\cdots ,s_n\right\} =\emptyset $ and $\left\{ r_1,\cdots ,r_m\right\}
\cup \left\{ s_1,\cdots ,s_n\right\} =\left\{ 1,2,\cdots ,N\right\} ,$ then
we obtain $\rho _{\left( \left\{ r_1,\cdots ,r_m\right\} ,\left\{ s_1,\cdots
,s_n\right\} \right) }.$ In addition, the $\rho _{kl}$ in Eq.($1)$ obviously
is a special case of two `bunches' containing only a single particle, or use
our symbol, $\rho _{kl}\equiv \rho _{\left( k,l\right) }$.

By the above ways, let $\left( k_i\right) _m\equiv \left\{ k_1,\cdots
,k_m\right\} $ and $\left( l_j\right) _n\equiv \left\{ l_1,\cdots
,l_n\right\} $ both be two subsets of $\left\{ 1,2,\cdots ,N\right\} ,$
where $m+n\leq N,\;1\leq k_1<k_2<\cdots <k_m\leq N,\;\;1\leq l_1<l_2<\cdots
<l_n\leq N$, and $\left\{ k_1,\cdots ,k_m\right\} \cap $ $\left\{ l_1,\cdots
,l_n\right\} =\emptyset ,$ then $\rho _{\left( \left( k_i\right) _m,\;\left(
l_j\right) _n\right) }$ is an entangled state between two bunches $\left(
\left\{ \mid i_{k_1}>_{k_1},\cdots ,\mid i_{k_m}>_{k_m}\right\} ,\left\{
\mid j_{l_1}>_{l_1},\cdots ,\mid j_{l_n}>_{l_n}\right\} \right) $. The set
of all possible $E_f\left[ \rho _{\left( \left( k_i\right) _m,\;\left(
l_j\right) _n\right) }\right] $ (notice that there are repeats in $\left\{
\rho _{\left( \left( k_i\right) _m,\;\left( l_j\right) _n\right) }\right\}
), $ is yet a description of character of the entanglement among the $m+n$
particles $\mid i_{k_1}>_{k_1},\cdots ,\mid i_{k_m}>_{k_m},\mid
j_{l_1}>_{l_1},\cdots ,\mid j_{l_n}>_{l_n}$ in the N-partite qubit state $%
\rho $, independent what happens to the remaining $N-m-n$ particles$.$

Secondly, as a special example we consider the GHZ state $\rho
_{GHZ_N}\equiv \mid $GHZ$_N><$GHZ$_N\mid (N\geq 3)$. By using of the above $%
\left( k_i\right) _m$ and $\left( l_j\right) _n,$we have the following
results

\begin{eqnarray}
\rho _{\left( \left( k_i\right) _m,\;\left( l_j\right) _n\right) }\text{ is
disentanled, }E_f\left[ \rho _{\left( \left( k_i\right) _m,\;\left(
l_j\right) _n\right) }\right] &=&0,\text{ for }2\leq m+n<N  \nonumber \\
\rho _{\left( \left( k_i\right) _m,\;\left( l_j\right) _n\right) }\text{ is
maximally entanled, }E_f\left[ \rho _{\left( \left( k_i\right) _m,\;\left(
l_j\right) _n\right) }\right] &=&1,\text{ for }m+n=N
\end{eqnarray}
The proof only is a straightforward calculation by Eqs.(17) and (18). This
result shows fully the character of $\rho _{GHZ_N}$, i.e. only when all N
particles are divided into two parts (every particle must be in one and only
one of them), the entanglement between this two parts does not vanish, and
it is maximal. Therefore in view of this, for N$\geq 3$ the result that all $%
E_f\left[ \left( \rho _{GHZ_N}\right) _{kl}\right] =0$ is not at all
surprising.

Other interesting example is ($w$ is a given integer$,1\leq w<N)$ 
\begin{equation}
\phi _{\left( N,w\right) }^{+}=\frac 1{\sqrt{2}}\left( \mid 0>_1\otimes
\cdots \otimes \mid 0>_w\otimes \mid 1>_{w+1}\otimes \cdots \otimes \mid
1>_N+\mid 1>_1\otimes \cdots \otimes \mid 1>_w\otimes \mid 0>_{w+1}\otimes
\cdots \otimes \mid 0>_N\right)
\end{equation}
it like to the Bell state $\varphi ^{+}\equiv \frac 1{\sqrt{2}}$ $\left(
\mid 0>_a\otimes \mid 1>_b+\mid 1>_a\otimes \mid 0>_b\right) .$ For $%
B_{\left( N,w\right) }^{+}\equiv \mid \phi _{\left( N,w\right) }^{+}><\phi
_{\left( N,w\right) }^{+}\mid ,$ it is easily verified that 
\begin{eqnarray}
\left( B_{\left( N,w\right) }^{+}\right) _{\left( \left\{ k_1,\cdots
,k_m\right\} ,\left\{ l_1,\cdots ,l_n\right\} \right) }\text{ is
disentanled, }E_f\left[ \left( B_{\left( N,w\right) }^{+}\right) _{\left(
\left\{ k_1,\cdots ,k_m\right\} ,\left\{ l_1,\cdots ,l_n\right\} \right)
}\right] &=&0,\text{ for }2\leq m+n<N  \nonumber \\
\left( B_{\left( N,w\right) }^{+}\right) _{\left( \left\{ k_1,\cdots
,k_m\right\} ,\left\{ l_1,\cdots ,l_n\right\} \right) }\text{ is entangled, }%
E_f\left[ \left( B_{\left( N,w\right) }^{+}\right) _{\left( \left\{
k_1,\cdots ,k_m\right\} ,\left\{ l_1,\cdots ,l_n\right\} \right) }\right]
&>&0,\text{ for }m+n=N
\end{eqnarray}
More generally, if $N\leq M$ and$\;\left( r_k\right) _N\equiv \left\{
r_1,\cdots ,r_N\right\} $ is a subset containing $N$ elements in the set $%
\left\{ 1,2,\cdots ,M\right\} ,$ we can construct $\phi _{\left( \left(
r_k\right) _N,\;w\right) }^{+}$ and $B_{\left( \left( r_k\right)
_N\;,w\right) }^{+},$ according to Eq.$(21)$ for the set $\left\{ \mid
i_k>_{r_k}\right\} (k=1,\cdots ,N)$ of N particles$.$ We define the state 
\begin{equation}
\mid \Psi _{\left( M,N,w\right) }>\equiv \mid \phi _{\left( \left(
r_k\right) _N,\;w\right) }^{+}>\otimes \mid 0\cdots 0>_{rest}
\end{equation}
and the mixed state

\begin{equation}
B_{\left( M,N,w\right) }^{+}\equiv \sum_{\text{all possible }\left(
r_k\right) _N\subset \left\{ 1,2,\cdots ,M\right\} }x_{\left( r_k\right)
_N}\mid \Psi _{\left( M,N,w\right) }><\Psi _{\left( M,N,w\right) }\mid
\end{equation}
where the real numbers $x_{\left( r_k\right) _N}$ obey 0%
\mbox{$<$}%
$x_{\left( r_k\right) _N}\leq 1$ and $\sum\limits_{\text{all possible }%
\left( r_k\right) _N\subset \left\{ 1,2,\cdots ,M\right\} }x_{\left(
r_k\right) _N}=1$, then Eq.(22) still holds for $B_{\left( M,N,w\right)
}^{+}.$ The action of $B_{\left( M,N,w\right) }^{+}$ is somewhat like to an
`entanglement molecule'[3].

Sum up, in a multipartite qubit state $\rho $ there is a new kind of
entanglement, i.e. the entanglement between two bunches of particles,
independent what happens to the remaining particles, which can be described
by the special bipartite reduced density operators $\rho _{\left( \left(
k_i\right) _m,\;\left( l_j\right) _n\right) },$ and the measure of
entanglement can be represented by $E_f\left[ \rho _{\left( \left(
k_i\right) _m,\;\left( l_j\right) _n\right) }\right] .$

\end{document}